\documentclass[pra, twocolumn, floatfix,superscriptaddress]{revtex4}
\usepackage{graphicx}
\usepackage{epstopdf}
\usepackage{amsmath, amsfonts, amssymb, bm,color}

\def\ifa{Institute of Applied Physics, Academiei str. 5, MD-2028 Chi\c{s}in\u{a}u, Moldova}

\def\uniX{Department of Physics, Friedrich-Alexander-Universit\"{a}t Erlangen-N\"{u}rnberg, 91058 Erlangen, Germany}
\begin{document}
\title{Multiphonon quantum dynamics in cavity optomechanical systems}
\author{Mihai A. Macovei }
\email{mihai.macovei@ifa.md}
\affiliation{\ifa}

\author{Adriana P\'{a}lffy}
\email{adriana.palffy-buss@fau.de}
\affiliation{\uniX}

\date{\today}
\begin{abstract}
The multiphonon quantum dynamics in laser-pumped cavity optomechanical samples containing a vibrating mirror is investigated.
Especially, we focus on dispersive interaction regimes where the externally applied coherent field frequency detuning from the 
optical resonator frequency is not equal to the mirror's oscillating frequency or to its multiples. As a result, for moderately strong 
couplings among the involved subsystems, the quantum dynamics of this complex system is described by multiphonon absorption 
or emission processes, respectively. Particularly, we demonstrate efficient ways to monitor the phonon quantum dynamics via 
photon detection. The possibility to extract the relevant sample parameters, for instance, the coupling strength between the 
mechanical mirror and the electromagnetic field, is also discussed. 
\end{abstract}
\maketitle

\section{Introduction}
Cavity optomechanical systems that couple electromagnetic field radiation with nanomechanical or micromechanical motion of a vibrating component have 
proven their potential  in numerous different applications,  e.g., optical networks, quantum memories, quantum metrology, quantum amplifiers, quantum 
sensing or gravimetry \cite{walls_milb,optm1,qsens,brag,optm,natc,grav}. Many of these applications, whether advancing fundamental quantum physics 
or technological in nature, rely on the fact that the photon-phonon coupling renders possible the quantum cooling of quantized motion 
\cite{cool,cool1,cool2,cool3,cool4,cool5}. Cavity optomechanical systems also allow  the creation and control of quantum macroscopic Schr\"{o}dinger cat 
states in cavities with a moving mirror \cite{schr1,schr2}. Furthermore,  the output of an externally pumped cavity optomechanical system shows a clear 
evidence of  electromagnetically induced transparency phenomenon, called in this case optomechanically induced transparency \cite{eit1,eit2,eit3}. Based 
on this effect, a photon switch effect was demonstrated \cite{eits}. Transferred to a different frequency regime by a proper optomechanical interface, 
optomagnetically induced transparency may help to control  light-matter interaction of x-rays via optical photons  \cite{eitx}. 

Generally speaking, optomechanical systems may be viewed as pumped optical resonators containing Kerr-like nonlinear elements \cite{ker1}. As a 
consequence, the cavity optomechanical samples exhibit bi- or multi-stability, multiple photon blockades, and various types of entanglement or squeezing 
phenomena \cite{ker1,ker2,marq,mma,mblock,ker3,ker4,ker5}. The photon blockade effect in optomechanical systems, i.e. preventing  multiple photons 
from entering the cavity at the same time due to strong photon-photon interactions, was considered from theory side in  Ref.~\cite{phblk1}. 
Cross-Kerr interactions between the optical and phonon modes lead to generation of mechanical cat states \cite{mecat}, for instance. Since photons 
typically do not mutually interact, enhanced photon-photon interactions in these systems would be of great interest in quantum computation and 
quantum information processing  \cite{luk1,luk2,chuang}. Furthermore, in the single-photon strong-coupling regime and  good-cavity limit, the cavity 
response shows several resolved resonances at multiples of the mechanical frequency, respectively \cite{phblk2}. This effect can be used for instance to 
measure the mechanical frequency if other involved parameters are known. Also, strong laser driving in an optomechanical setup was 
analyzed in \cite{stlas}, where a transition from sub-Poissonian to super-Poissonian photon statistics was determined.

Most of the above mentioned results were obtained under the condition that the externally applied coherent field frequency detuning from the optical 
resonator frequency is equal or close to the mirror's oscillating frequency. In contrast, here we shall focus on the different regime when the above 
mentioned resonance condition is not met. The laser frequency detuning from the cavity one is considered unequal to the frequency of the mirror's 
vibrations or to its multiples. In addition, we consider the mechanical frequency is larger than the detuning and than the corresponding decay rates in 
the sample, and is commensurable with but larger than the coupling strength among the interacting subsystems. We demonstrate that under these 
circumstances, the quantum dynamics of the oscillating mirror has a multi-phonon nature in the sense that the quantities describing its evolution in 
steady-state are characterized by absorption or emission of many mechanical oscillation quanta. We show that in the proposed setup the corresponding 
phonon dynamics, proper to the mechanical part, follows that of the photon one. The calculation of the  mean-phonon as well as  mean-photon numbers 
demonstrated that detecting the leaking photons from the optical cavity one can monitor the vibrations of the moving subsystem. Correspondingly, the 
quantum nature of these processes is established through the second-order phonon-phonon or photon-photon correlation functions. Furthermore, our 
results show that the sample's parameters like the phonon-photon coupling strengths can be extracted from the multi-peak structure of the mean-photon 
number quantum dynamics. Finally, the parameter range needed to observe this behavior is close to those for photon blockade 
effect in optomechanical systems \cite{phblk1,phblk2}, and  within reach of experiments \cite{exp1,exp2,exp3}.

This paper is organized as follows. In Sec.~\ref{theo} we describe the analytical approach and the system of interest, while in Sec.~\ref{numres} we 
present and analyse the obtained numerical results. The article concludes with a summary and outlook in Sec.~\ref{sum}.

\section{Analytical approach \label{theo}}
We describe our sample using the master equation approach under Born-Markov approximations where the coupling of the relevant degrees of freedom 
to their environmental counterparts is weak, whereas the photon and phonon memory effects are negligible \cite{walls_milb,reww}. Appropriate unitary 
transformations performed further will allow us to follow the multi-phonon quantum dynamics of the mechanical part or the corresponding photon dynamics 
as a function of the ratio of the coupling strength $g$ over mechanical oscillation frequency $\omega$. This way one can distinguish the corresponding 
quantum dynamics of the photon-phonon subsystems, respectively, when single- or many-phonons are involved. 

The master equation describing a standard laser pumped cavity optomechanical setup, in the Born-Markov approximations and in a frame rotating 
at the external laser field frequency $\omega_{L}$ is given by
\begin{eqnarray}
\frac{d}{dt} \rho + \frac{i}{\hbar}[H,\rho] &=& -\frac{\kappa_{a}}{2}[a^{\dagger},a\rho] - \frac{\kappa_{b}}{2}(1+\bar n)[b^{\dagger},b\rho] \nonumber \\
&-& \frac{\kappa_{b}}{2}\bar n [b,b^{\dagger}\rho] + H.c.,
\label{MeqG}
\end{eqnarray}
where $\rho$ is the density matrix and the Hamiltonian $H$ is given by the expression
\begin{eqnarray}
H = \hbar\Delta a^{\dagger}a + \hbar \omega b^{\dagger}b + \hbar\epsilon (a^{\dagger} + a) + \hbar g a^{\dagger}a(b+b^{\dagger}).
\label{HG}
\end{eqnarray}
The coherent evolution of the examined system is described by the second term of the left-side part of the master equation (\ref{MeqG}). The damping effects of 
the involved photon and phonon subsystems are characterized by the right-side part of this equation,  with $\kappa_{a}$ and $\kappa_{b}$ being the corresponding 
photon or phonon damping rates, respectively. Here, $\bar n=(\exp[\hbar \omega/(k_{B}T)]-1)^{-1}$ is the mean-phonon number due to the thermal bath environment 
at temperature $T$ and at the vibration frequency $\omega$ of the mechanical resonator, while $k_{B}$ is the Bolzmann's constant. $a^{\dagger}(b^{\dagger})$ 
is the creation operator of a photon (phonon), whereas $a(b)$ is the corresponding photon (phonon) annihilation operator, respectively, satisfying the standard 
bosonic commutation relations $[a,a^{\dagger}]=1$, $[b,b^{\dagger}]=1$, and $[b^{\dagger},b^{\dagger}]=[b,b]=0$, $[a^{\dagger},a^{\dagger}]=[a,a]=0$. 
The first and the second components from the Hamiltonian (\ref{HG}) account for the free energies of the optical and mechanical resonators, respectively, with 
$\Delta=\omega_{c}-\omega_{L}$ being the detuning of the optical cavity frequency $\omega_{c}$ from the laser one. The third term in (\ref{HG}) describes 
the laser pumping effects of the optical resonator's mode with $\epsilon$ being the corresponding amplitude. The last term accounts for the interaction among 
the optical and mechanical motion degrees of freedom characterized by the coupling strength $g$.

In the following, we consider that $\omega >  g > \epsilon$, and do not yet impose any conditions on  $\kappa_{a}$ and $\kappa_{b}$. Note that when the 
optomechanical coupling is comparable to or larger than the optical decay rate and the mechanical frequency, the steady state of the mechanical oscillator 
can develop a nonclassical strongly negative Wigner density \cite{nwign}. So, we proceed to perform a unitary transformation
\begin{eqnarray}
U=e^{\chi a^{\dagger}a(b-b^{\dagger})},
\label{UT}
\end{eqnarray}
in the master equation (\ref{MeqG}) leading to the following new bosonic operators
\begin{eqnarray}
\bar b^{\dagger} = \chi \bar a^{\dagger} \bar a + b^{\dagger}, ~~~ \bar b = \chi \bar a^{\dagger} \bar a + b,
\label{tb}
\end{eqnarray}
and
\begin{eqnarray}
\bar a^{\dagger} = a^{\dagger}e^{\chi(\bar b - \bar b^{\dagger})}, ~~~ \bar a = a e^{-\chi(\bar b - \bar b^{\dagger})}.
\label{ta}
\end{eqnarray}
Upon the unitary transformation (\ref{UT}), the master equation (\ref{MeqG}) takes the form:
\begin{eqnarray}
\frac{d}{dt} \bar \rho &+& \frac{i}{\hbar}[\bar H,\bar \rho] = -\frac{\kappa_{a}}{2}[\bar a^{\dagger}e^{-\chi(\bar b - \bar b^{\dagger})},\bar a e^{\chi(\bar b 
- \bar b^{\dagger})}\bar \rho] \nonumber \\
&-& \frac{\kappa_{b}}{2}(1+\bar n)[(\bar b^{\dagger}-\chi\bar a^{\dagger}\bar a ),(\bar b - \chi \bar a^{\dagger}\bar a)\bar \rho] \nonumber \\
&-& \frac{\kappa_{b}}{2}\bar n [(\bar b - \chi\bar a^{\dagger}\bar a), (\bar b^{\dagger}-\chi\bar a^{\dagger}\bar a)\bar \rho] + H.c.,
\label{MeqT}
\end{eqnarray}
where 
\begin{eqnarray}
\bar H &=& \hbar\Delta \bar a^{\dagger}\bar a + \hbar \omega \bar b^{\dagger}\bar b + \hbar\epsilon \bigl(\bar a e^{\chi(\bar b - \bar b^{\dagger})} 
+ \bar a^{\dagger}e^{-\chi(\bar b - \bar b^{\dagger})}\bigr) 
\nonumber \\
&-& \omega\chi^{2}(\bar a^{\dagger}\bar a)^{2},
\label{HT}
\end{eqnarray}
with
\begin{eqnarray}
\chi=\frac{g}{\omega}.
\end{eqnarray}
The last term in the Hamiltonian (\ref{HT}) describes the vibration-induced Kerr-like nonlinearity effects. Furthermore, the master equation (\ref{MeqT}) exhibits 
resonance conditions if $\Delta = \pm k\omega$, $\{k=1,2,\cdots\}$. In what follows, we shall focus on the case  when $\Delta \not=\pm k\omega$, but rather 
$\omega>\Delta$.

For the parameter regime of interest, we can proceed by expanding the exponents in Eqs.~(\ref{MeqT}-\ref{HT}) in the Taylor series using the small parameter $\chi \ll 1$, 
\begin{eqnarray}
e^{\pm \chi(\bar b - \bar b^{\dagger})} = \sum^{\infty}_{n=0}\frac{(\pm\chi)^{n}}{n!}(\bar b - \bar b^{\dagger})^{n}.
\label{dsc}
\end{eqnarray}
Further, using the bosonic operator identity
\begin{eqnarray}
(A+ B)^{n} = \sum^{n}_{k}\frac{n!}{k!(\frac{n-k}{2})!}\bigl(-C/2\bigr)^{\frac{n-k}{2}}\sum^{k}_{r=0}\binom{k}{r}
 A^{r}B^{k - r}, \nonumber \\
\label{BI}
\end{eqnarray}
where $[A,B]=C$ and $[A,C]=[B,C]=0$, whereas $k$ is odd for an odd $n$ and even for an even $n$, respectively, one can simplify the expression 
(\ref{dsc}) depending on the assumed approximations. Performing a unitary transformation 
\begin{eqnarray}
V=\exp[i(\Delta \bar a^{\dagger}\bar a + \omega \bar b^{\dagger}\bar b)t],
\label{UT2}
\end{eqnarray}
in the master equation (\ref{MeqT}) and avoiding any resonances in the system, i.e. $\Delta \pm k\omega \not=0$, $\{k=1,2,\cdots\}$, one can neglect 
then all time-dependent terms $\propto e^{\pm ik\omega t}$ in the master equation. This approximation additionally requires that  
$\omega \gg \{\kappa_{a},\kappa_{b}\}$. We keep those terms oscillating at $e^{\pm i\Delta t}$, which accounts to further assuming that $\Delta/\omega \ll1$. 
As a result, the exponent expression (\ref{dsc}), which enters in the Hamiltonian (\ref{HT}), for instance, takes the following form in this case:
\begin{eqnarray}
e^{\pm \chi(\bar b - \bar b^{\dagger})} = \sum^{\infty}_{n=0}\sum^{n}_{m=0}\frac{(-1)^{m}\chi^{2n}}{(m!)^{2}(n-m)!}
\frac{\bar b^{m}\bar b^{\dagger m}}{2^{n-m}}. \label{Hexp}
\end{eqnarray}

Respectively, the exponent expression entering in the optical resonator's damping in Eq.~(\ref{MeqT}), i.e. 
\begin{eqnarray*}
&{}&e^{\chi(\bar b - \bar b^{\dagger})}\bar a\bar \rho \bar a^{\dagger}e^{-\chi(\bar b - \bar b^{\dagger})} = \nonumber \\
&{}& \sum^{\infty}_{n_{1}n_{2}=0}\frac{(-\chi)^{n_{1}}\chi^{n_{2}}}{n_{1}!n_{2}!}\bigl(\bar b^{\dagger} - 
\bar b\bigr)^{n_{1}}\bar a\bar \rho \bar a^{\dagger}\bigl(\bar b^{\dagger} - \bar b\bigr)^{n_{2}}, 
\end{eqnarray*}
acquires  the form:
\begin{widetext}
\begin{eqnarray}
e^{\chi(\bar b - \bar b^{\dagger})}\bar a\bar \rho \bar a^{\dagger}e^{-\chi(\bar b - \bar b^{\dagger})} &=&\sum^{\infty}_{n_{1}n_{2}=0}
\sum^{n_{1}}_{k_{1}}\sum^{n_{2}}_{k_{2}}\frac{(-\chi)^{n_{1}}\chi^{n_{2}}}{k_{1}!k_{2}!}\frac{(1/2)^{(n_{1}-k_{1})/2}(1/2)^{(n_{2}-k_{2})/2}}
{[(n_{1}-k_{1})/2]![(n_{2}-k_{2})/2]!}\sum^{k_{1}}_{r_{1}=0}\sum^{k_{2}}_{r_{2}=0}(-1)^{r_{1}+r_{2}}\binom{k_{1}}{r_{1}}\binom{k_{2}}{r_{2}}
\nonumber \\
&\times& \bar b^{r_{1}}\bar b^{\dagger k_{1}-r_{1}}\bar a\bar \rho \bar a^{\dagger}\bar b^{r_{2}}\bar b^{\dagger k_{2}-r_{2}}e^{i(k_{1}-2r_{1})\omega t}
e^{i(k_{2}-2r_{2})\omega t}, \label{crtMq}
\end{eqnarray}
\end{widetext}
where, again, $k_{i}$, $\{i=1,2\}$, is odd for an odd $n_{i}$ and even for an even $n_{i}$, respectively. It is easy to observe that the above expression is 
time-independent if
\begin{equation}
k_{1} - 2r_{1} + k_{2} - 2r_{2}=0. \label{kf}
\end{equation}
Expressions (\ref{Hexp}-\ref{kf}) have to be introduced in the master equation (\ref{MeqT}) and one can already recognize the multiphonon 
nature of the cavity optomechanical dynamics in the off-resonance situation considered here. 

Once we have arrived at a time-independent master equation, we can obtain the corresponding equation for photon-phonon distribution function, namely, 
$P_{n_{1}n_{2},m_{1}m_{2}}$=$\langle n_{1},m_{1}|\bar \rho|m_{2},n_{2}\rangle$, where indices $n(m)$, $\{n,m=0,1,2,\cdots\}$, 
refers to photon(phonon) subsystem, respectively, with $|n(m)\rangle$ being the corresponding Fock state. Actually, that equation is diagonal with 
respect to phonon degrees of freedom [see Eq.~\eqref{MeqT} with expressions \eqref{Hexp}-\eqref{kf} and \eqref{ap1}]. Therefore, the final distribution 
function will be represented as $ P_{n_{1}n_{2},mm}= \langle n_{1},m|\bar \rho|m,n_{2}\rangle$. Furthermore, the $N$-phonon processes are described 
by terms proportional to $\chi^{2N}$, see Appendix A. In the presence of corresponding damping effects, we can calculate the photon-phonon distribution 
function $P_{n_{1}n_{2},mm}$ numerically in steady-state.

From the master equation (\ref{MeqT}), modified based on relations (\ref{Hexp}-\ref{kf}), one obtains an infinite number of equations for the 
photon-phonon distribution function $P_{n_{1}n_{2},mm}$. In order to solve the infinite system of equations for $P_{n_{1}n_{2},mm}$, we 
truncate it at a certain maximum value $\{n=n_{max},m=m_{max}\}$ so that a further increase of its value, i.e. $\{n_{max},m_{max}\}$, does 
not modify the obtained results if other involved parameters are being fixed. Thus, using the operator relations (\ref{tb}-\ref{ta}) and keeping the 
time-independent terms only, the optical resonator's steady-state mean quanta number can be expressed as
\begin{eqnarray}
\langle a^{\dagger}a\rangle = \sum^{n_{max}}_{n=0}\sum^{m_{max}}_{m=0}nP_{nm}, \label{ampm}
\end{eqnarray}
while the mechanical vibrational  mean-phonon number is
\begin{eqnarray}
\langle b^{\dagger}b\rangle = \sum^{n_{max}}_{n=0}\sum^{m_{max}}_{m=0}(m+\chi^{2}n^{2})P_{nm}. \label{bmpm}
\end{eqnarray}
Here $P_{nm} \equiv P_{nn,mm} = \langle n,m|\bar \rho|m,n\rangle$ with
\begin{eqnarray}
\sum^{n_{max}}_{n=0}\sum^{m_{max}}_{m=0}P_{nm}=1. \label{nrm}
\end{eqnarray}
The second-order photon-photon correlation function \cite{glauber} is defined as
\begin{eqnarray}
g^{(2)}_{a}(0)=\bigl(1/\langle a^{\dagger}a\rangle^{2}\bigr)\sum^{n_{max}}_{n=0}\sum^{m_{max}}_{m=0}n(n-1)P_{nm}, 
\label{gndoi}
\end{eqnarray}
whereas the second-order phonon-phonon correlation function is calculated via the expression:
\begin{eqnarray}
g^{(2)}_{b}(0)&=&\bigl(1/\langle b^{\dagger}b\rangle^{2}\bigr)\sum^{n_{max}}_{n=0}\sum^{m_{max}}_{m=0}\bigl(m(m-1) 
+ 4\chi^{2}mn^{2} \nonumber \\
&+& \chi^{4}n^{4}\bigr)P_{nm}. 
\label{gmdoi}
\end{eqnarray}
We note here that in a cavity optomechanical setup, second-order phonon-phonon correlation functions are measured experimentally, 
as demonstrated e.g. in Ref.~\cite{exp4}. 
\begin{figure}[t]
\includegraphics[width = 4.3cm]{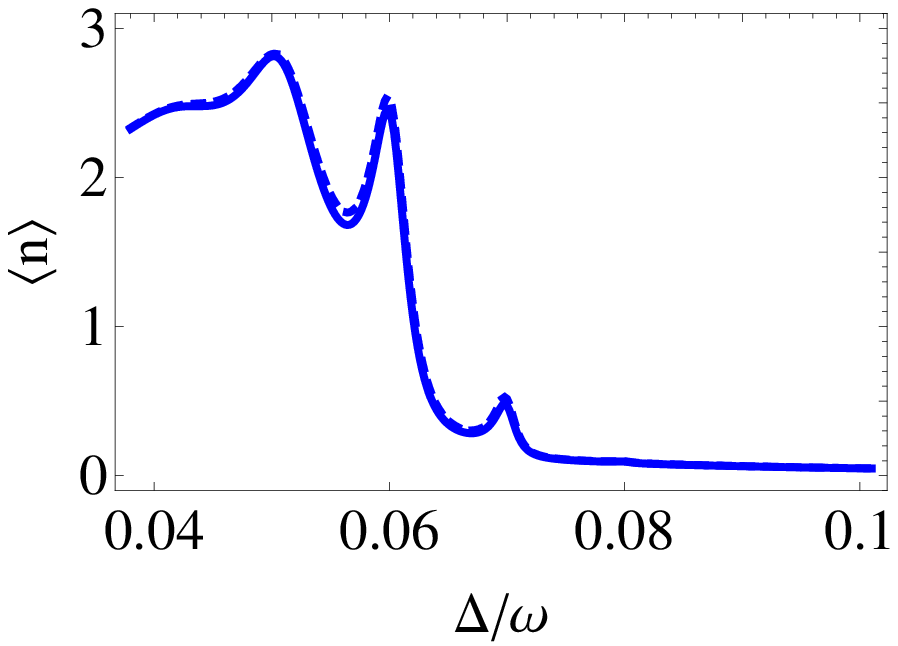}
\hspace{-0.2cm}
\includegraphics[width = 4.33cm]{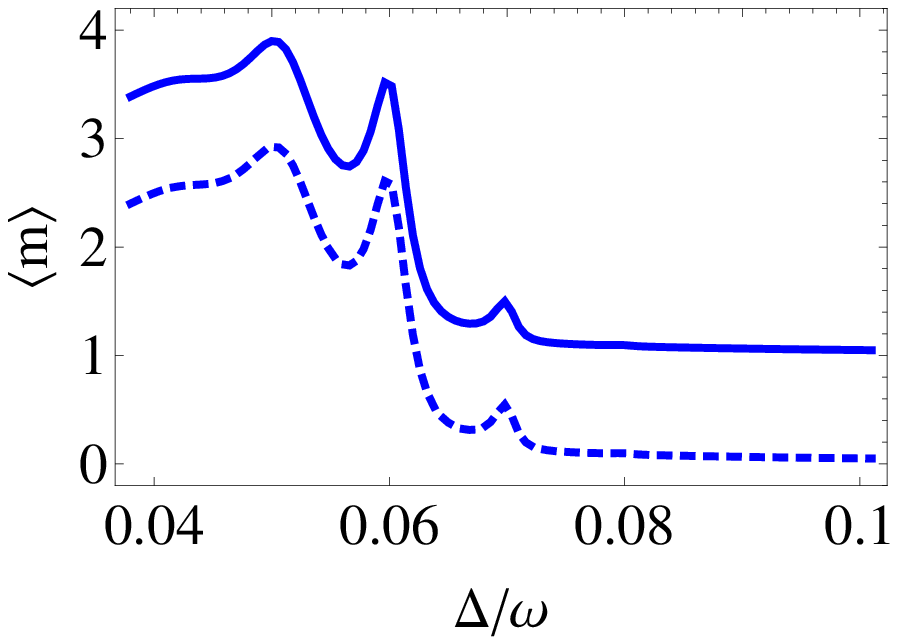}
\begin{picture}(0,0)
\put(-20,85){(a)}
\put(105,85){(b)}
\end{picture}
\caption{\label{fig-1} 
(a) The steady-state optical cavity mean photon number $\langle n\rangle \equiv \langle a^{\dagger}a\rangle$ as well as (b) the corresponding 
mechanical resonator mean phonon number $\langle m\rangle \equiv \langle b^{\dagger}b\rangle$  as a function of  $\Delta/\omega$. 
Here $\chi=0.1$, $\epsilon/\omega=0.02$, $\kappa_{a}/\omega = 2\cdot 10^{-3}$ and $\kappa_{b}/\omega = 2\cdot 10^{-5}$, 
and only single-phonon processes were included. Also, $\bar n=1$ for solid lines, while $\bar n=0$ for short-dashed lines, respectively.}
\end{figure}
\begin{figure}[b]
\includegraphics[width = 4.27cm]{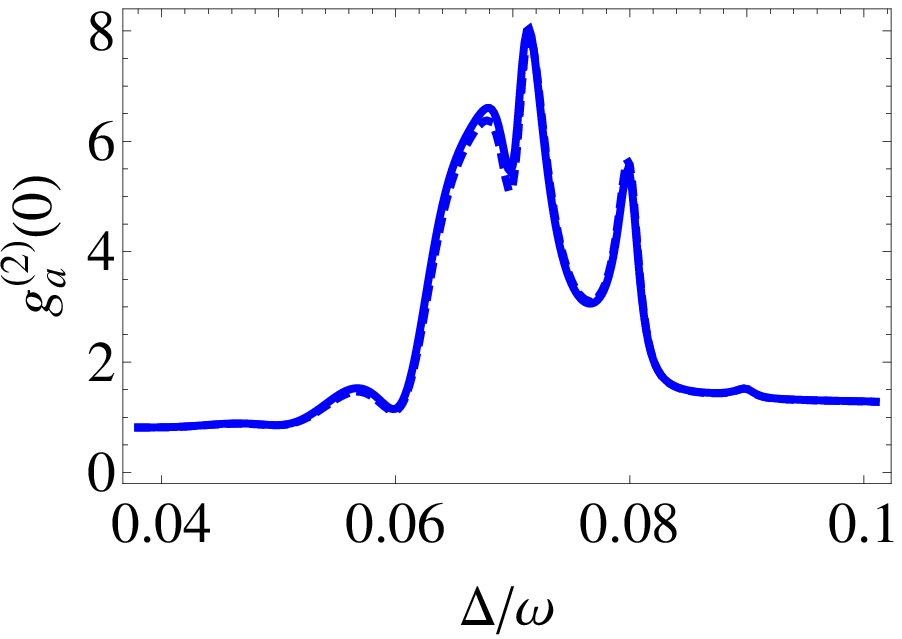}
\hspace{-0.2cm}
\includegraphics[width = 4.35cm]{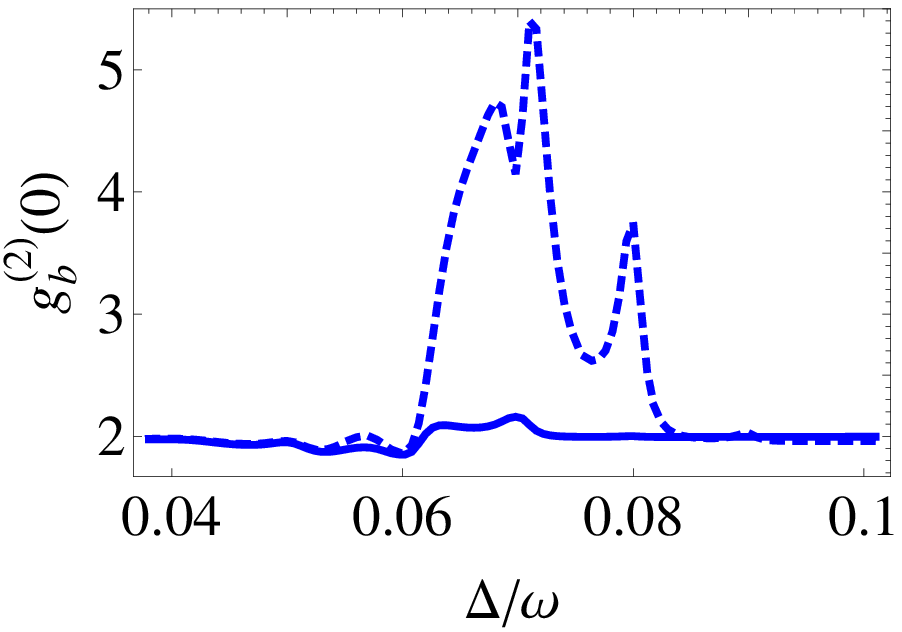}
\begin{picture}(0,0)
\put(-20,85){(a)}
\put(105,83){(b)}
\end{picture}
\caption{\label{fig-2} 
The steady-state second-order correlation function for (a) photons, $g^{(2)}_{a}(0)$, and (b) for phonons, $g^{(2)}_{b}(0)$, as a function 
of $\Delta/\omega$. The solid lines are for $\bar n=1$, while the short-dashed one for $\bar n=0$, respectively. 
Other parameters are as in Fig.~\ref{fig-1}.}
\end{figure}

\section{Numerical Results and Discussion \label{numres}}
In Fig.~\ref{fig-1} we present the numerical results of the steady-state mean photon and phonon numbers for a generic system with parameters within the 
considered regime with $\chi=0.1$, $\epsilon/\omega=0.02$, $\kappa_{a}/\omega = 2\cdot 10^{-3}$ and $\kappa_{b}/\omega = 2\cdot 10^{-5}$. 
These parameters are accessible experimentally. For instance,  a similar set with $\omega \sim 1\rm{MHz}$, $g \sim 0.3\rm{MHz}$, $\kappa_{a}\sim 0.2\rm{MHz}$ 
and $\kappa_{b}\sim 150\rm{Hz}$ was reported in Ref.~\cite{exp1}. For our case, also a good-cavity regime with   $\omega \gg \kappa_{a}$ would be of interest, 
since  a too strong photon loss washes out the predicted features. Furthermore, we considered here only one-phonon process, which corresponds to keeping terms 
up to $\chi^{2}$ in Eq.~(\ref{MeqT}), respectively. Few interesting features can be observed: (i) In these parameter ranges, the steady-state phonon dynamics is 
quite sensitive on external temperature variations, (ii) the photon dynamics follows that of phonon one (see Appendix \ref{appB}), and vice versa, though with a 
different magnitude, and (iii) the peak frequency intervals equal the Kerr-like nonlinearity $\omega \chi^{2}$, see Hamiltonian (\ref{HT}). The latter result can be 
intuitively understood from the free-photon Hamiltonian plus the Kerr-like contribution, i.e. 
$\bar H_{0}= \hbar \bar a^{\dagger}\bar a(\Delta - \omega \chi^{2}\bar a^{\dagger}\bar a)$, from where it follows that the effective detuning vanishes at 
$\Delta_{k}=k\omega \chi^{2}$ or $|\Delta_{k}-\Delta_{k \pm 1}|=\omega \chi^{2} \equiv g^{2}/\omega$. This reveals that the induced Kerr nonlinearity 
shifts the frequency of the corresponding photon Fock states, which can be detected by scanning the frequency of the cavity photons. Thus, the mean-phonon 
number dynamics can be extracted via detection of the mean-photon number's quantum dynamics, while the coupling constant $g$ can be estimated if the 
mechanical motion frequency $\omega$ is known (or vice versa). 
\begin{figure}[t]
\includegraphics[width = 4.2cm]{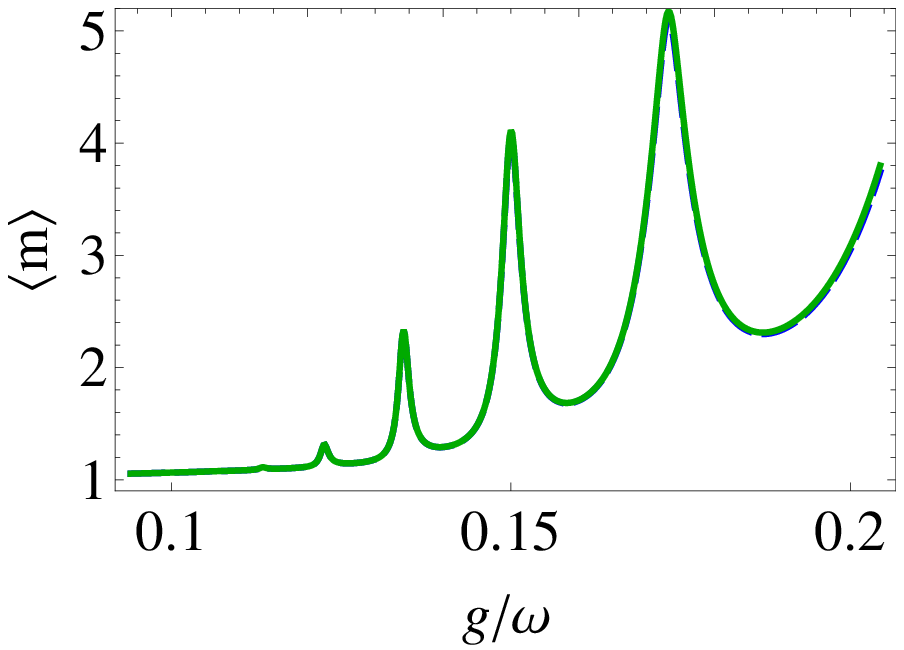}
\hspace{-0.15cm}
\includegraphics[width = 4.39cm]{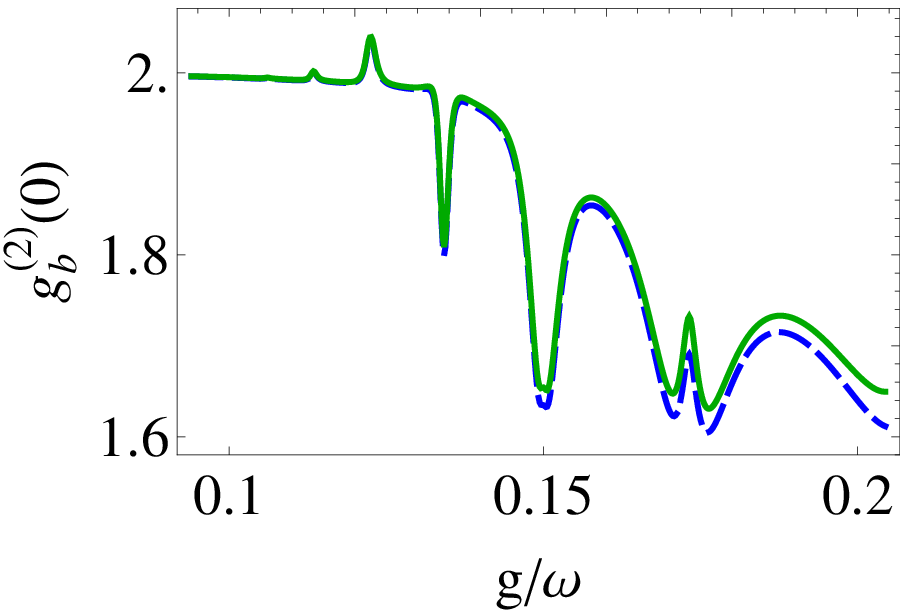}
\begin{picture}(0,0)
\put(-20,85){(a)}
\put(105,83){(b)}
\end{picture}
\caption{\label{fig-3} 
The steady-state behaviors of (a) the mean-phonon number $\langle m\rangle$ as well as (b) the second-order phonon-phonon correlation function 
$g^{(2)}_{b}(0)$ as a function of $\chi=g/\omega$. The dashed blue line is plotted for single-phonon whereas the green solid one for three-phonon 
processes, respectively. Here $\Delta/\omega=0.09$ and $\bar n=1$ while other parameters are as in Fig.~(\ref{fig-1}).}
\end{figure}

Figure \ref{fig-2} shows the corresponding second-order correlation functions. The photon statistics changes from sub-Poissonian ($g^{(2)}_{a}(0) < 1$) 
for $\Delta/\omega \lesssim 0.05$, to super-Poissonian ($g^{(2)}_{a}(0) \ge 2$) in the interval 0.06-0.08 and to quasi-coherent features 
($g^{(2)}_{a}(0) \gtrapprox 1$ )  for $\Delta/\omega > 0.08$. On the other hand, the phonon statistics exhibits quasi-thermal to super-Poissonian 
properties. Also in this case, the curves are characterized by a multi-peak structure. For smaller and negative $\Delta/\omega$, the lines in Figs.~(\ref{fig-1}) 
and (\ref{fig-2}) remain flat.  Notice that in the absence of vibrations, i.e. when $g=0$,  the cavity mean-photon number and its correlation function are given 
by the following expressions in the steady-state:
\begin{eqnarray*}
\langle a^{\dagger}a\rangle &=& \epsilon^{2}/\bigl(\Delta^{2}+(\kappa_{a}/2)^{2}\bigr), ~~{\rm and} \nonumber \\
\langle a^{\dagger 2}a^{2}\rangle &=& \epsilon^{4}/\bigl(\Delta^{2}+(\kappa_{a}/2)^{2}\bigr)^{2}, 
\end{eqnarray*}
leading to $g^{(2)}_{a}(0) \equiv \langle a^{\dagger 2}a^{2}\rangle/\langle a^{\dagger}a\rangle^{2} =1$.
For an independent boson mode in thermal equilibrium with its surrounding thermal bath we obtain 
\begin{eqnarray*}
\langle b^{\dagger}b\rangle =\bar n, ~~{\rm and}~~ \langle b^{\dagger 2}b^{2}\rangle=2\bar n^{2}, 
\end{eqnarray*}
with $g^{(2)}_{b}(0) \equiv \langle b^{\dagger 2}b^{2}\rangle/\langle b^{\dagger}b\rangle^{2} =2$. 

In Fig.~\ref{fig-3} we investigate the mean phonon number and its statistics as a function of $\chi=g/\omega$ for a fixed ratio $\Delta/\omega=0.09$. 
As can be seen in Fig.~\ref{fig-1}, for the particular coupling strength $g$ chosen ($\chi=0.1$) the mean-quanta numbers are quite small at this detuning.
Figure \ref{fig-3}  compares the case of single-phonon processes, i.e. keeping terms up to $\chi^{2}$ in Eq.~(\ref{MeqT}), with the case of three-phonon 
processes, for which we consider all terms up to $\chi^{6}$ in Eq.~(\ref{MeqT}). As a first observation, we notice that the mean-phonon number increases 
with the ratio $g/\omega$. The peaks are slightly higher for three-phonon processes, but the differences are barely noticeable in Fig.~\ref{fig-3}a. 
Furthermore, the maxima occur at $\chi_{k}$=$\sqrt{\Delta/(k\omega)}$, which for $\Delta/\omega=0.09$, give k=6, 5, 4 and 3. Also, the steady-state 
mean phonon number is direct proportional to the mean thermal phonon number $\bar n$ due to the environmental reservoir, see Appendix \ref{appB}. 
The second-order phonon-phonon correlation function changes accordingly and exhibits quasi-thermal features for $\bar n=1$. In this case, single-phonon 
and three-phonon processes become distinguished at higher ratios of $g/\omega$, see Fig.~\ref{fig-3}(b). Thus, multiphonon processes occur for 
higher ratios of $g/\omega$. It can be shown that also in this case the mean photon number dynamics is following the mean phonon number dynamics, 
albeit with a smaller scaling. Finally, we have also calculated higher-order phonon correlation functions \cite{multix,multiy}, i.e. 
$g^{(k)}_{b}(0)=\langle b^{\dagger k}b^{k}\rangle/\langle b^{\dagger}b\rangle^{k}$ as a function of $\Delta/\omega$ with $k \in \{3,4\}$. Our 
numerical results show that these curves closely follow the ones  of $g^{(2)}_{b}(0)$, however with a larger magnitude. For instance, the peak at 
$\Delta/\omega=0.07$ appears for the third-order (fourth order) correlation function at the same position, but with a value of approx. 7.6 (39).

Our results confirm that the presence of a vibrating mirror changes significantly both the photon's as well as the phonon quantum dynamics of an 
off-resonant pumped cavity optomechanical system. Moreover, the parameter range required to observe this behavior is 
within  experimental reach \cite{exp1,exp2,exp3} and is close to those for photon blockade effect predicted in optomechanical systems 
\cite{phblk1,phblk2}. Finally, the analytical approaches proper to cavity optomecanics with a moving mirror equally apply to other related 
samples, like e.g., hybrid metal-dielectric cavities \cite{hdcav}, plasmon-excitonic polaritons \cite{pep}, superconducting qubits and quantum 
circuits \cite{qcirc} or other types of nanomechanical resonators \cite{nmr}, respectively, rendering our  developed analytical approach relevant also for 
these systems. 

\section{Conclusions \label{sum}}
We have investigated a cavity optomechanical setup where the detuning of the external coherent electromagnetic field frequency from the optical 
resonator's one is different from the frequency (or its multiples) of one of the vibrating mirror. As a result, the quantum dynamics of this complex 
system is accompanied by multiple phonon absorption and emission processes. We have computed the mean-quanta numbers and the corresponding 
second-order correlation functions and described their properties. Particularly, we have found that the inter-peak frequency intervals observed 
in the quantum dynamics of the photon and phonon subsystem as a function of detuning equal the vibration-induced Kerr-like non-linearity. The 
photon mean-number dynamics follows that of the mean-phonon one which is convenient for monitoring the phonon quantum dynamics by photon 
detection. The corresponding second-order correlation functions also exhibit a multi-peak structure and are completely different from the fixed-mirror 
case. The photon-photon correlation function may exhibit sub-Poissonian to super-Poissonian photon statistics. The corresponding correlation function
for phonons lies within quasi-thermal to thermal phonon statistics, characterized by super-Poissonian features. Furthermore, the second order phonon 
correlation function can be used to distinguish among the one- and few-phonon processes for stronger coupling strengths among the involved interacting 
subsystems. For a more detailed investigation of non-classical signatures, future work could focus on the Wigner function for representative 
parameter sets, or on so-called bundle correlation functions, $g^{(2)}_2(\tau)=\langle b^{\dagger 2}(0)b^{\dagger 2}(\tau)b^{2}(\tau)b^{2}(0)\rangle/\langle
(b^{\dagger2}b^2)(0)\rangle\langle (b^{\dagger2}b^2)(\tau)\rangle$ \cite{multix,multiy}, which could reveal highly correlated phonon behaviour.

\acknowledgments
MM acknowledges the financial support by the Moldavian National Agency for Research and Development, grant No. 20.80009.5007.07. 
AP gratefully acknowledges support from the Heisenberg Program of the Deutsche Forschungsgemeinschaft (DFG).

\appendix
\section{Expanding the exponents up to $\chi^{4}$ \label{appA}}
In order to demonstrate the multi-phonon nature of the combined quantum dynamics, let us expand the exponents in the expression 
(\ref{crtMq}), i.e. $e^{\chi(\bar b - \bar b^{\dagger})}\bar a\bar \rho \bar a^{\dagger}e^{-\chi(\bar b - \bar b^{\dagger})}$, up to 
$\chi^{4}$ and keep the time-independent terms only, that is,
\begin{widetext}
\begin{eqnarray}
e^{\chi(\bar b - \bar b^{\dagger})}\bar a\bar \rho \bar a^{\dagger}e^{-\chi(\bar b - \bar b^{\dagger})} = \bar a\bar \rho \bar a^{\dagger} 
+ \biggl(\bar a\bar \rho \bar a^{\dagger}\bigl\{\frac{\chi^{2}}{2!}(1-2\bar b\bar b^{\dagger}) +  \frac{\chi^{4}}{4!}(6\bar b^{2}\bar b^{\dagger 2} 
- 12\bar b\bar b^{\dagger}+ 3)\bigr\} + H.c.\biggr) + \chi^{2}(1+\chi^{2})\biggl (\bar b \bar a\bar \rho \bar a^{\dagger}\bar b^{\dagger} \nonumber \\
+ \bar b^{\dagger} \bar a\bar \rho \bar a^{\dagger}\bar b \biggr) - \frac{\chi^{4}}{2}\biggl(\bar b \bar a\bar \rho \bar a^{\dagger}\bar b \bar b^{\dagger 2} 
+ \bar b^{\dagger} \bar a\bar \rho \bar a^{\dagger}\bar b^{2}\bar b^{\dagger} + H.c.\biggr) + \frac{\chi^{4}}{4}\biggl(\bar b^{2}\bar a\bar \rho 
\bar a^{\dagger}\bar b^{\dagger 2} + (1-2\bar b\bar b^{\dagger})\bar a\bar \rho \bar a^{\dagger}(1-2\bar b\bar b^{\dagger}) + \bar b^{\dagger 2}
\bar a\bar \rho \bar a^{\dagger}\bar b^{2} \biggr). \nonumber \\
\label{ap1}
\end{eqnarray}
\end{widetext}
When considering the photon-phonon distribution function $P_{n_{1}n_{2},mm}$, then from (\ref{ap1}) one can observe that terms proportional 
at least to $\chi^{2}$ contribute to single-phonon processes, whereas  those proportional to $\chi^{4}$ - to two-phonon processes, respectively.
For instance, the last term from the first line of expression (\ref{ap1}) accounts for single-phonon processes where the phonon number changes by 
$\pm 1$, i.e. $m \pm 1$. On the other side, the last term from (\ref{ap1}) describes two-phonon processes where the phonon number modifies by 
$\pm 2$, $m \pm 2$. Thus, the expression (\ref{ap1}), expanded up to $\chi^{4}$ terms, describes simultaneously single- and two-phonon processes, 
respectively. Also, single-phonon processes are influenced by the second-order ones, i.e. there is a contribution to single-phonon processes coming 
from $\chi^{4}$ terms.

Generalizing now, one can state that $N$-phonon processes are described by terms proportional to $\chi^{2N}$. There is a contribution to $N$-phonon 
processes which arises from higher order terms proportional to $\chi^{2(N+1)}$. This way, depending on a certain power of $\chi$, one can investigate 
the multiphonon quantum dynamics of the cavity optomechanical system.

\section{Relationship among the optical and phonon modes \label{appB}}
In what follows, we shall give details on how the mean cavity photon and mean phonon numbers are interconnected in the dispersive interaction 
regime investigated here. The master equation (\ref{MeqT}), containing terms up to $\chi^{2}$ for simplicity and in the considered approximations, 
reads 
\begin{eqnarray}
\frac{d}{dt}\bar \rho(t) &+& \frac{i}{\hbar}[\bar H,\bar \rho] =  \nonumber \\
&-& \frac{\kappa_{a}}{2}\biggl(\bar a^{\dagger}\bar a \bar \rho - e^{\chi(\bar b - \bar b^{\dagger})}\bar a\bar \rho \bar a^{\dagger}
e^{-\chi(\bar b - \bar b^{\dagger})} \biggr)\nonumber 
\end{eqnarray}
\begin{eqnarray}
&-& \frac{\kappa_{b}}{2}(1+\bar n)\biggl([\bar b^{\dagger},\bar b\bar \rho] + \chi^{2}[\bar a^{\dagger}\bar a,\bar a^{\dagger}\bar a\bar \rho]\biggr) 
\nonumber \\
&-& \frac{\kappa_{b}}{2}\bar n\biggl( [\bar b,\bar b^{\dagger}\bar \rho] + \chi^{2}[\bar a^{\dagger}\bar a,\bar a^{\dagger}\bar a\bar \rho]\biggr)  + H.c.,
\nonumber \\ 
\label{app2}
\end{eqnarray}
where 
\begin{eqnarray}
\bar H &=& \hbar\Delta \bar a^{\dagger}\bar a + \hbar \omega \bar b^{\dagger}\bar b + \hbar\epsilon(\bar a + \bar a^{\dagger})\bigl\{1-\chi^{2}(\bar b^{\dagger}\bar b
+ \bar b\bar b^{\dagger})/2\bigr\} \nonumber \\
&-& \omega\chi^{2}(\bar a^{\dagger}\bar a)^{2}. \label{app3}
\end{eqnarray}
The exponent term from the first line of the damping part of the above master equation can be obtained from (\ref{ap1}) while setting $\chi^{4}\to 0$.
Then, in the steady-state, from the master equation (\ref{app2}) one can easily show that
\begin{eqnarray}
\langle \bar b^{\dagger}\bar b\rangle = \bar n + \chi^{2}\frac{\kappa_{a}}{\kappa_{b}}\langle \bar a^{\dagger}\bar a\rangle,
\label{app4}
\end{eqnarray}
while from Eqs.~(\ref{tb}), one finally arrives at the relationship among the steady-state mean phonon and photon numbers, namely,
\begin{eqnarray}
\langle b^{\dagger}b\rangle &=& \langle \bar b^{\dagger}\bar b\rangle + \chi^{2}\langle \bar a^{\dagger 2}\bar a^{2}\rangle \nonumber  \\
&=& \bar n + \chi^{2}\frac{\kappa_{a}}{\kappa_{b}}\langle \bar a^{\dagger}\bar a\rangle + \chi^{2}\langle \bar a^{\dagger 2}\bar a^{2}\rangle.
\label{app5}
\end{eqnarray}
One can observe here that there is an almost  linear dependence between mean phonon and cavity photon numbers, respectively, since the last term 
may give little contribution as long as $\chi \ll 1$. In addition,  we have assumed that $\kappa_{a}/\kappa_{b} \gg 1$ . Thus, detecting the cavity photons 
one can estimate the mean phonon number and vice versa.


\end{document}